\documentclass[twocolumn,amsmath,amssymb,aps,pra,longbibliography]{revtex4-2}
\usepackage{dcolumn}
\usepackage{bm}
\usepackage{mathtools, cases}
\usepackage{float}
\usepackage{graphicx}
\usepackage[unicode]{hyperref}
\hypersetup{colorlinks=true,linkcolor=blue,citecolor=blue,filecolor=black,urlcolor=blue}
\newcommand{\bi}[1]{Fig.~\ref{fig:#1}} 
\newcommand{\e}[1]{Eq.~(\ref{eq:#1})}
\begin{document}
\title{Self-consistent autocorrelation of a disordered Kuramoto model in the asynchronous state}
\author{Yagmur Kati$^{1,*}$, Jonas Ranft$^{3}$, Benjamin Lindner$^{1,2}$}
\affiliation{
${\ }^1$ Department of Physics, Humboldt Universität zu Berlin, Newtonstr 15, 12489 Berlin, Germany\\
${\ }^2$ Bernstein Center for Computational Neuroscience, Haus 2, Philippstr 13, 10115 Berlin, Germany\\
${\ }^3$ Institut de Biologie de l’ENS, Ecole Normale Supérieure, CNRS, Inserm, Université PSL, 46 rue d’Ulm, 75005 Paris, France}
\email{kati@physik.hu-berlin.de}
\date{\today}
\begin{abstract}
The Kuramoto model has provided deep insights into synchronization phenomena and remains an important paradigm to study the dynamics of coupled oscillators. Yet, despite its success, the asynchronous regime in the Kuramoto model has received limited attention. 
Here, we adapt and enhance the mean-field approach originally proposed by Stiller and Radons [Phys.~Rev.~E 58 (1998)] to study the asynchronous state in the Kuramoto model with 
a finite number of oscillators and 
with disordered connectivity. By employing an iterative stochastic mean field (IMF) approximation, the complex $N$-oscillator system can effectively be reduced to a one-dimensional dynamics, both 
for homogeneous and heterogeneous networks. This method allows us to investigate the power spectra of individual oscillators as well as of the multiplicative ``network noise'' in the Kuramoto model in the asynchronous regime. By taking into account 
the finite system size and disorder in the connectivity, our findings become relevant for the dynamics of coupled oscillators that appear in the context of biological or technical systems.  
\end{abstract}

\maketitle

\section{Introduction}

The synchronization of oscillators has been a focal point of interest for multiple scientific disciplines, with applications spanning from neuroscience to power grids \cite{Kur84,PikRos01,AceBon05}. While synchronized states and their properties have been explored in depth, the asynchronous states, which are also prevalent in many systems \cite{Poulet08,Harris11,Vreeswijk96,Renart10,Helias14,Ostojic2014,VanLin18}, remain a less explored area of study. 
In particular, the classical Kuramoto model has provided a foundation for understanding synchronization in a system (or network) of coupled phase oscillators \cite{kuramoto1975self,AceBon05}. However, real-world networks often exhibit more intricacies as e.g.~disorder in both the natural frequencies of individual oscillators and the network topology (interaction strengths), which can lead to more nuanced behaviors that the basic model might lack.
While the effects of different types of heterogeneity in the network have begun to be addressed in earlier studies, these mostly focused on the  characterization of different dynamical regimes with at least partial synchrony (see, e.g., \cite{Bansal19}). Here, we are more interested in the \emph{asynchronous} state, which has received relatively little attention in comparison. In particular for neuronal networks, synchronous states can often be considered pathological, whereas the asynchronous state corresponds frequently to the default \cite{Jingwen20}. To advance our understanding of network dynamics in such asynchronous states, we therefore aim to deliver here a proper characterization of such states in the paradigmatic Kuramoto model.

Important characteristics of the asynchronous state are the fluctuation statistics of the single oscillators 
as well as of the effective drive of individual oscillators due to couplings. The open problem for theory is to find approximations for correlation functions or power spectra of these observables. 
It is interesting to see how the fluctuation statistics depends on the disorder of the oscillator frequencies and the connectivity. 
If the dynamics of the single oscillators is more complicated, e.g. given by a multidimensional system of differential equations for each oscillator, then the above questions can only be addressed by numerical simulations of the network. However, such simulations of a large network of oscillators can be computationally expensive and do not offer much mechanistic understanding of the fluctuation statistics. In this context, the iterative mean field (IMF) approach, which effectively reduces the $N$-oscillator system to a one-dimensional self-consistent representation, emerges as a powerful tool. It has been applied to recurrent network models of rate units \cite{SomCri88,AljSte15,MasOst17}, of integrate-and-fire neurons \cite{LerSte06,DumWie14,PenVel18}, of rotator units \cite{VanLin18,RanLin22,RanLin23}, and to disordered chains of Ising spins \cite{EisOpp92}. For a Kuramoto model with random connectivity, the stochastic mean field theory 
that underlies the iterative approach was worked out  by Stiller and Radons \cite{stiller} but applied only to the relaxation of the order parameter, but not to the stationary fluctuations. The results of this simplified description, as we will illustrate here, agree well with those of the full network dynamics for heterogeneous Kuramoto networks with frequency and/or connectivity disorder. Specifically, we aim here to analyze the effects of randomness in interactions and leverage the IMF approximation to this end. We focus our investigation on the spectra of the single oscillators and of the network noise, presenting a fresh perspective on the asynchronous dynamics of oscillator networks. In addition to expanding the theoretical framework, our investigation also serves to fill the gaps left by previous research on the heterogeneous Kuramoto model. Prior attempts have approached the numerical integration of 1D dynamics with iteration primarily to compute the time-dependent order parameter \cite{stiller,Dai92}. However, the main focus here is the temporal correlation statistics of the stationary fluctuations.

Our paper is organized as follows. We start by presenting the network model and the stochastic iterative mean field  (IMF) method used to compute the spectral statistics, followed by a detailed comparison of analytical and numerical results. We then take a detailed look at the network noise power spectra properties, ending with a summary and discussion of our key findings. 

\section{The Model}
We investigate a system described by the heterogeneous Kuramoto model, a paradigm for weakly coupled oscillators. Each oscillator in this network is characterized by the following dynamics of its phase $\theta_\ell$ ($\ell=1,\dots,N$): 
\begin{equation}\label{eq:dottheta}
    \dot{\theta}_\ell(t)= \omega_\ell + \sum_{m=1}^N {K_{\ell m}} \sin{(\theta_m(t)-\theta_\ell(t))} +   \xi_\ell(t).
\end{equation} 
Here, $\omega_\ell$ are the natural frequencies of the oscillators, $K_{\ell m}$ are the coupling coefficients between the oscillators, and $\xi_\ell$ are independent noise processes. The $\omega_\ell$ are taken to be Gaussian distributed with
\begin{align}
&\omega_\ell=G_\ell \sigma_\omega, \quad\langle\omega_\ell\rangle = 0, \quad  \langle\omega_\ell \omega_m\rangle=\sigma_\omega^2 \delta_{\ell m} \label{eq:omega} 
\end{align}
where $G_\ell$ (for $\ell=1,\dots,N$) represents a Gaussian array of $N$ independent numbers with vanishing mean and unit variance. 
The coupling coefficients $K_{\ell m}$ are also taken to be independent Gaussian numbers with
\begin{align}
& {K_{\ell m}}= \frac{K}{N}   + \frac{k G_{\ell m}}{\sqrt{N}},\quad \langle K_{\ell m} \rangle=\frac{K}{N},\nonumber \\
&\langle K_{\ell m} K_{\ell' n} \rangle= \delta_{\ell{\ell'}} \delta_{m{n}} \frac{{k}^2}{N} + \frac{K^2}{N^2} \label{eq:kij}  
\end{align}
where $G_{\ell m}$ (for $\ell,m=1,\dots,N$) denotes a $N\times N$ Gaussian matrix with mutually independent entries of standard deviation 1 and mean 0.
Note that in our setting 
$K_{\ell m}$ and $K_{m\ell}$ are uncorrelated. 
We consider the  $\xi_\ell(t)$ to be Gaussian white noise with $\langle \xi_\ell (t) \xi_m(t') \rangle=2D \delta(t-t') \delta_{\ell m}$, i.e.~uncorrelated between individual oscillators, with noise intensity $D$. We initialize the system with all oscillator phases $\theta_\ell$ drawn independently from a uniform distribution in $[0,2\pi]$. 

\begin{figure}[tp]
\centering
\includegraphics[width=1\linewidth]{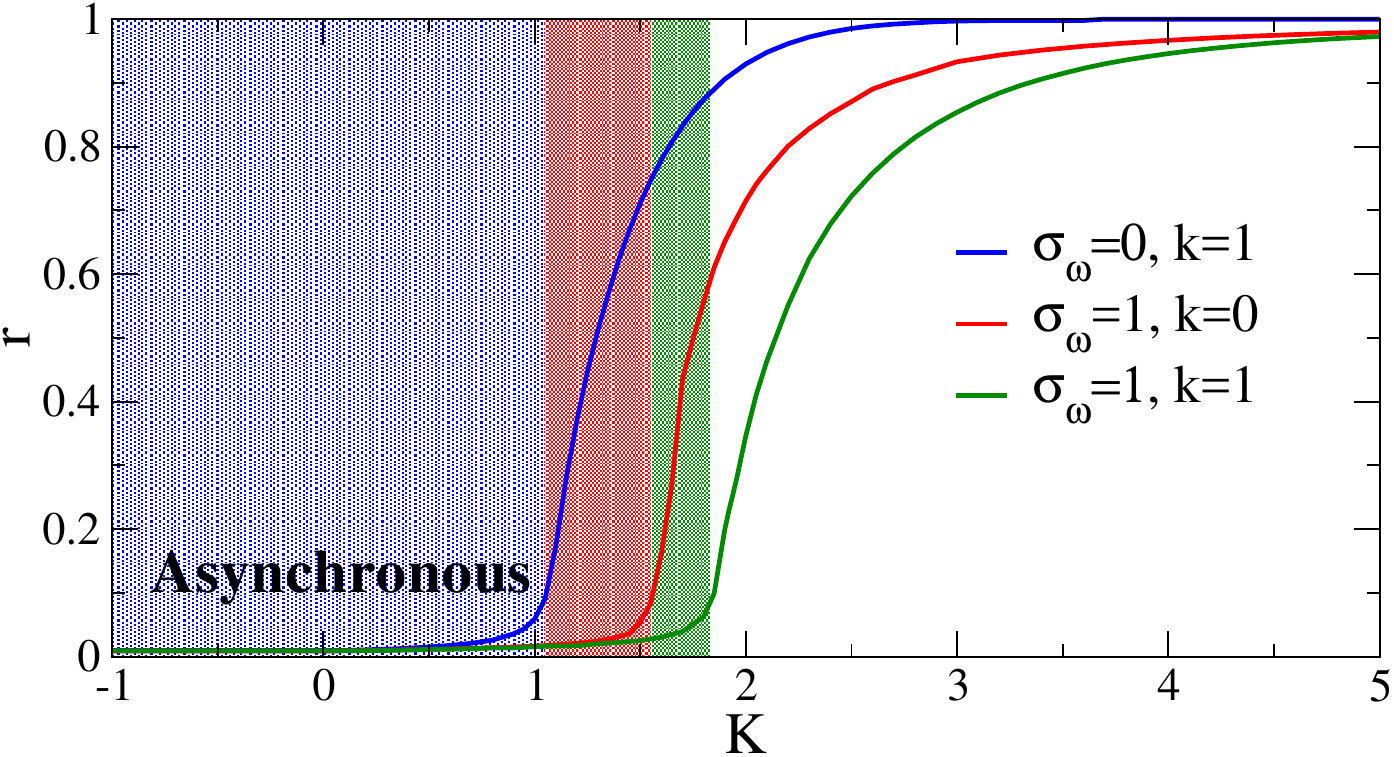}
\caption{\textbf{Synchronous and asynchronous regimes of the model}. Variation of the order parameter $r$ as a function of coupling strength $K$ (solid lines). There are three scenarios: disorder in connectivity (blue), disorder in frequencies (red), and concurrent disorder in both connectivity and frequencies (green). This paper exclusively focuses on the asynchronous regime (shaded areas). Parameters used: $N=10^4, T=10^3, R=20$.}
\label{fig:fig1}
\end{figure}

In our study for completely deterministic network oscillators ($D=0$), we used the Runge-Kutta method to explicitly integrate Eqs.~\eqref{eq:dottheta}; in the stochastic case ($D>0$), we used the Euler-Maruyama method. In both cases, we refer to the integration of Eqs.~\eqref{eq:dottheta} in the following as network dynamics (ND) method. We discard a transient simulation time of $t_d=10^3$ and use a high relative tolerance of the Runge-Kutta method from $10^{-8}$ down to $10^{-14}$ (for $D=0$). The time window of further computations (e.g. Fourier transforms) is denoted by $T$ and chosen depending on parameters from $10^3$ to $10^5$ (see figure captions). We also performed simulations with simple Euler-Maruyama method (for $D=0$ and $D>0$) with time step $dt=0.01$ and obtained (for $D=0$) results that were quantitatively very close to those of the Runge-Kutta method.

In order to distinguish synchronous and asynchronous regimes in the model, it is useful to inspect the order parameter of the Kuramoto model, $r$, defined by
\begin{equation}
r e^{i\psi} = \frac{1}{N} \sum_{\ell=1}^N e^{i\theta_\ell}.
\end{equation}
For a finite network, both the phase $\psi(t)$ and the order parameter $r$ will show fluctuations that can be reduced by temporal averaging. In \bi{fig1}, we present the time-averaged order parameter $r$ obtained by averaging over the time window $T$ after the transient $t_d$ is discarded. Subsequently, this averaged result is further processed by averaging over $R$ disorder realizations of connectivities and frequencies.
The figure displays the order parameter $r$ versus the average coupling constant $K$ for three different cases: oscillator-frequency disorder with $\sigma_{\omega}=1$ and no disorder in the couplings (blue line), connectivity disorder with $k=1$ and no disorder in the natural frequencies (red line), and a combination of both types of disorder. 
As expected, the stronger disorder in the system leads to a larger range of coupling strengths for which the system is in an asynchronous state, i.e., has a small order parameter $r\ll 1$. Generally, asynchronous states in complex systems occur when the average coupling constant $K$ is below a critical coupling $K_c$. In the asynchronous regime, the order parameter scales as $1/\sqrt{N}$ \cite{PhysRevE.88.032126}. This can be motivated by looking at the squared order parameter's expected value:
\begin{equation}
   \!\langle r^2 \rangle\! \! = \!\left\langle \!\! \frac{1}{N^2} \sum_\ell \sum_{\ell'}\! e^{i\theta_\ell} e^{i\theta_{\ell'}} \!\!\right\rangle \! =\! \frac{1}{N} \!+ (1-\frac{1}{N}) \langle e^{i(\theta_1-\theta_2)} \rangle \!\approx \frac{1}{N}.
\end{equation}
In the last but one step, we simplify the sum by choosing two distinct oscillators (all pairs of distinct oscillators are statistically equivalent). 
In the last step, we take into account that the oscillators are asynchronous, i.e., $\theta_1, \theta_2$ are statistically independent and their phases are uniformly distributed.

As a measure of the fluctuations in the asynchronous state, we analyze the power spectrum $S_{x_\ell}$ of individual oscillators in networks 
with different types of disorder: in the natural frequencies, in the connectivities, and in both. An oscillator is represented here by its complex pointer
$x_\ell = e^{i\theta_\ell}$, allowing us to compute its power spectrum with
\begin{equation}\label{eq:sxl}
S_{x_\ell} = \lim_{T\rightarrow\infty}\frac{\langle \tilde{x}_\ell {\tilde{x}^*}_\ell \rangle}{T}, \qquad  \tilde{x_\ell}=\int_0^T dt e^{i\omega t}x_\ell(t).
\end{equation} 
We can also look at the power spectrum averaged over all oscillators:
\begin{equation}
S_{z}=\frac{1}{N}\sum_{i=1}^N S_{x_\ell },
\end{equation}
which in the asynchronous state, where individual oscillators are uncorrelated, is equivalent to the power spectrum of the observable
\begin{equation}
    z(t)=\frac{1}{\sqrt{N}} \sum_{\ell=1}^N x_\ell.
\end{equation}
For the numerical evaluation of the average of the power spectra in \e{sxl} we cannot perform the limit $T\to\infty$ but have to use a sufficiently large time window; depending on the parameters, we use here time windows of $T=10^4$ or $T=10^5$. The average is performed in three ways, depending on the specific measure considered.
For the single oscillator spectra, we average only over time, keeping the network connectivity and frequencies of the oscillators fixed.   
The time average is carried out by smoothing the raw spectrum over neighboring frequency bins such that the resulting coarse-grained frequency bin is $\Delta\omega=2\pi/100$ (the effective time window of $10^3$ is still large enough for an estimation of a power spectrum). In other situations, as indicated, we will also average over $R$ draws of the oscillator frequencies $\omega_\ell$ and/or the coupling coefficients $K_{\ell m}$.

We are interested in the dependence of these power spectra on the properties of the network, i.e. the disorder in the frequencies, in connectivities, and the dynamical noise. We study both large as well as small systems. 

\section{Derivation of Self-Consistent Equation}

Stiller and Radons have developed a stochastic mean-field theory \cite{stiller} that describes the dynamics of a single oscillator driven by a Gaussian network noise with self-consistent autocorrelation statistics. Their derivation is based on the method of generating functionals, and they applied their result to the problem of relaxation into the steady state using a method pioneered by Eissfeller and Opper \cite{EisOpp92}. Here we give an alternative, simplified derivation of the stochastic mean-field dynamics and apply the self-consistent iterative method to the problem of stationary power spectra of single oscillators and network noise. Note that a similar approach has been used for random networks of integrate-and-fire neurons~\cite{LerSte06,DumWie14,PenVel18}, pulse-coupled~\cite{VanLin18} and non-Kuramoto coupled oscillators~\cite{VanLin18,RanLin22,RanLin23}.

In our setting of the model, we deviate in two respects from the model considered by Stiller and Radons: i) the mean value of the interaction is not zero, i.e., we also include in this way the original Kuramoto model; ii) the interaction between oscillators is completely random (this is the special case with Stiller and Radon's parameter $\eta=0$). 
Furthermore, we include a finite size correction in our theory that is not present in Stiller and Radon's theory.

Starting from the network dynamics defined by \e{dottheta}, we rewrite the sine coupling using complex exponentials, which permits us to write the coupling term as a multiplicative network noise that is multiplied with a complex exponential of the driven phase variable: 
\begin{align}
   \dot{\theta_\ell}&= \omega_\ell+\xi_\ell +   \sum_{m=1}^N \frac{K_{\ell m}}{2i} \left(e^{i(\theta_m(t)-\theta_\ell(t))} - e^{-i(\theta_m(t)-\theta_\ell(t))} \right) \nonumber \\
  & = \omega_\ell+\xi_\ell +  \text{Im}\left(e^{-i\theta_\ell(t)} \zeta_\ell (t) \right). 
\end{align}
Here we introduced the network noise 
\begin{align}
 & \zeta_\ell (t)=\sum_m K_{\ell m} e^{i\theta_m}. \label{eq:zetaell}
 \end{align}
 For a large network in the asynchronous state, this noise comprises many independent stochastic processes, a superposition with Gaussian statistics by virtue of the central limit theorem. For the correlation function of this noise, we find
 \begin{align}
    &\left\langle \zeta_\ell ^*(t) \zeta_\ell (t') \right\rangle=  \sum_{m,n}  \left \langle  K_{\ell m} K_{\ell{n}} e^{i(\theta_m(t')-\theta_{n}(t))} \right\rangle \\
    &= \sum_{m,{n}} \left(\frac{{k}^2}{N} \delta_{m{n}}+\frac{K^2}{N^2} \right) \left\langle e^{i(\theta_m(t')-\theta_{n}(t))}\right\rangle,
 \end{align}
 where we have used that the connection strengths $K_{\ell m}$ and $K_{\ell{n}}$ are uncorrelated with the phases of the driving ($m$th and $n$th) oscillators, and thus the exponential term and the product of coupling strengths can be separately averaged, which also eliminates the explicit dependence on $\ell$ (the correlation statistics of the network noise is the same for all oscillators). Taking into account that different oscillators will be uncorrelated in the asynchronous state, $\langle e^{i(\theta_m(t')-\theta_{n}(t))}\rangle=\delta_{mn}\langle e^{i(\theta_m(t')-\theta_{m}(t))}\rangle$, we can carry out the sum and arrive at
  \begin{align}
   \left\langle \zeta ^*(t) \zeta (t') \right\rangle &=\left(k^2 +\frac{K^2}{N}\right)  \left\langle e^{i(\theta(t')-\theta(t))}  \right\rangle.   \label{eq:zeta1}  
\end{align} 
Importantly, on the right-hand side, we find essentially the autocorrelation function of the pointer $e^{i\theta(t)}$ of a single phase variable, but averaged over all oscillators; note that we therefore suppressed the indices on both sides. In Fourier space, the same relation reads 
\begin{equation}
S_\zeta(\omega)= (k^2 + K^2/N)S_z(\omega). 
\label{eq:spec_zeta}
\end{equation}
The network noise power spectrum is directly proportional to the power spectrum of the single oscillator's pointer,  averaged over all oscillators, which reflects the self-consistence between the activity of the single rotator and the fluctuation by which it is driven.

To summarize, in the large-$N$ limit, we can transform the $N$ oscillator problem \e{dottheta} into a single oscillator problem with self-consistent noise statistics:
\begin{align}
& \dot{\theta}=\sigma_\omega \xi_\omega +\sqrt{2D}\xi(t)+  \text{Im}\left(e^{-i\theta} \zeta(t) \right) \label{eq:thetadotmean}
\end{align}
where $\xi(t)$ is a Gaussian white noise with correlation function $\langle \xi(t) \xi(t') \rangle =\delta(t-t')$ and $\xi_\omega$ is a static noise obeying $\langle\xi_\omega(t) \xi_\omega(t') \rangle =1$. The real-valued Gaussian noise processes 
$\xi_\omega$ and $\xi(t)$ and the complex-valued network noise $\zeta(t)$ have all zero mean. While the statistics of $\xi_\omega$ and  $\xi(t)$ are known beforehand and can simply be simulated in \e{thetadotmean}, the statistics of the network noise has to be determined self-consistently from the driven phase variable via \e{zeta1} or its Fourier variant \e{spec_zeta}, which is in turn shaped by the network noise. The noise statistics can be determined by an iterative procedure detailed below. We note that the analytical approach of 
Ref.~\cite{VanLin18} cannot be applied in the current case because the equivalent network noise (or total input) $\text{Im}\left(e^{-i\theta} \zeta(t) \right)$ appearing in \e{thetadotmean} turns out to be non-Gaussian.

\begin{figure}[hbt!] 
\centering
\includegraphics[width=1\linewidth]{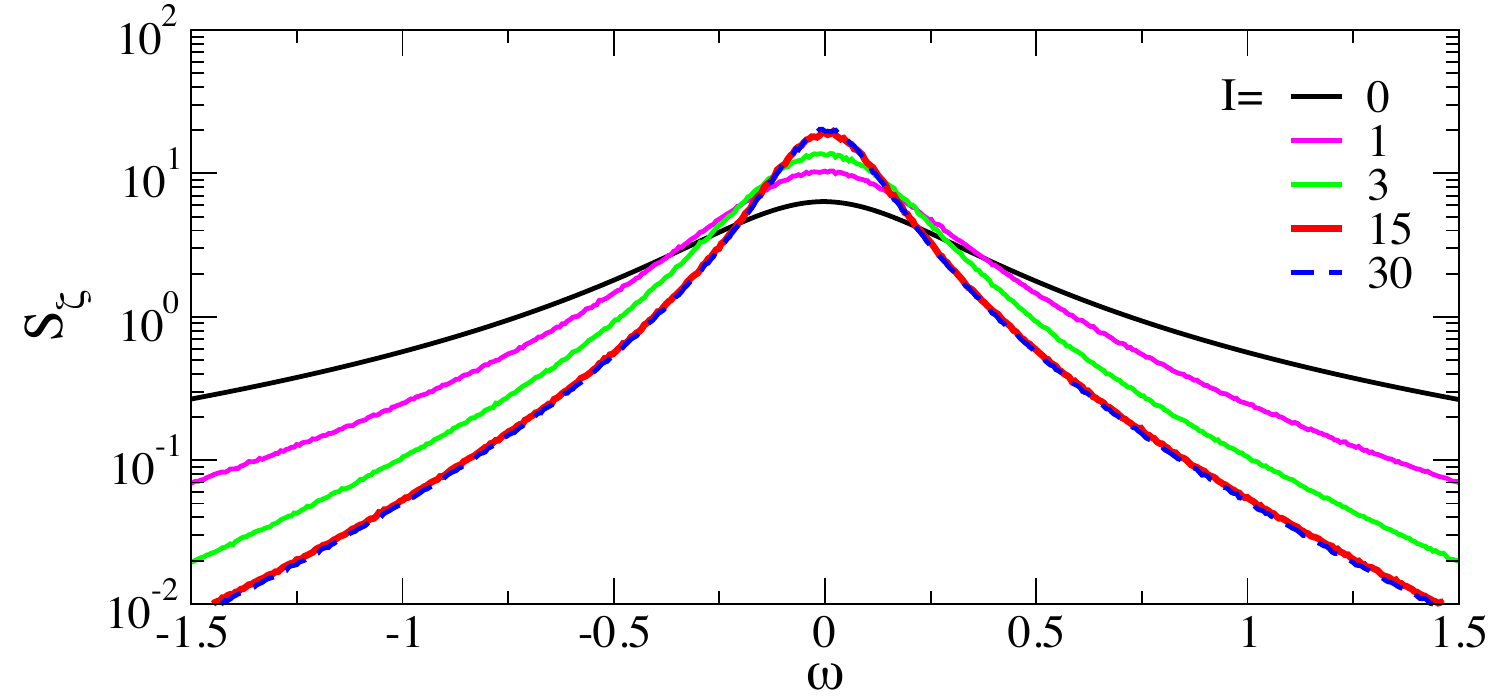}
\caption{\textbf{Convergence of the iterative mean field (IMF) approach for the self-consistent noise spectrum}. Noise spectra $S_\zeta$ labeled by the iterative step; we start with a Lorentzian spectrum i.e. $2/(\pi^2 + 20 \omega^2)$ (labeled as 0) and quickly approach a self-consistent shape (spectra for 15 and 30 steps agree within line thickness). Parameters: $k=1$, $K=0$, $\sigma_\omega=0$, $T=10^3$, $dt=10^{-3}$, with $N_{trial}=10^4$ number of realizations.}  
\label{fig:fig2}
\end{figure}

At the core of what we refer to in the following as \emph{iterative mean field} (IMF) method stands the insight that a Gaussian noise with a prescribed power spectrum can be easily produced by generating random numbers in the frequency domain and subsequent Fourier transformation \cite{BilShi90}. More specifically, if we draw Gaussian numbers $G_\ell, G'_\ell$ with zero mean and unit variance in each frequency bin (uncorrelated with each other and among the frequency bins), then 
\begin{equation}\label{eq:tildezeta}
 \tilde{\zeta}(\omega) = (G_\ell + i G'_\ell) \sqrt{\frac{S_\zeta(\omega) T}{2}}   
\end{equation}
constitutes the Fourier transform of a Gaussian noise with power spectrum $S_\zeta(\omega)$, the inverse Fourier transform of which will provide a sample of this surrogate noise process. 

In our iterative routine, we start with a Gaussian noise with a Lorentzian power spectrum $\gamma/[\pi(\omega^2 +\gamma^2)]$ 
and then solve \e{thetadotmean} for a number of realizations that we categorize as follows. For a given random value of the frequency $\sigma_\omega \xi_\omega$, we produce $M$ trials (realizations) by generating different realizations of the network noise. We repeat this $N$ times with $N$ different frequencies, such that the total number of trials is $N_\text{trial}=N\cdot M$. From the $N_\text{trial}$ trials, we compute the power spectrum of the pointer $e^{i\theta(t)}$ and determine the next (improved) version of the power spectrum of the network noise $S_{\zeta}(\omega)$ via \e{spec_zeta}. The latter is then used to obtain surrogate noise samples for the next iteration step. Note that the number of oscillators $N$ in the network enters twice in the IMF simulation scheme: Once by the amplitude of the network noise in \e{spec_zeta} (later used in \e{tildezeta}) and, secondly, by the number of frequencies drawn in each iterative step. Iterations are repeated until the spectra converge. After convergence has been reached, we can also obtain the spectrum of an individual oscillator with frequency $\omega_\ell=\sigma_\omega \xi_\omega$ by simulating \e{thetadotmean} with a fixed eigenfrequency and subsequent Fourier transformation of the 
dynamics of the pointer $e^{i\theta(t)}$.

In \bi{fig2}, we illustrate the effective convergence of $S_\zeta$ using the IMF method, beginning with a Lorentzian spectrum (black line) as the initial state. The method reaches stable convergence in this case at the 15th iteration (red line), which is illustrated by its agreement with the result for the 30th iteration (blue dashed 
line).  We furthermore validate the IMF approach by the systematic comparison  with numerical results from network dynamics (ND).

In the regime dominated by the intrinsic white noise $\xi_\ell$, the system's dynamics are described by $\dot{\theta} \approx \sqrt{2D}\xi(t)$ (we can neglect all the other terms on the right-hand side if the white noise is very strong). Under the Gaussian-white-noise assumption, the autocorrelation function, $C(t-t') = \langle e^{i[\theta(t) - \theta(t')]} \rangle$, reduces to $e^{-D |t - t'|}$, a special case of the Kubo oscillator \cite{Gar85} (the general Kubo oscillator is driven by a temporally correlated Gaussian noise).  
The power spectra, derived through the Fourier transform, are succinctly expressed as 
\begin{equation}\label{eq:d}
S_z = 2D / (D^2 + \omega^2),    
\end{equation}
providing a simple testable limit case. 

\section{Results}\label{sec:results}

\begin{figure}
\centering
\includegraphics[width=1\linewidth]{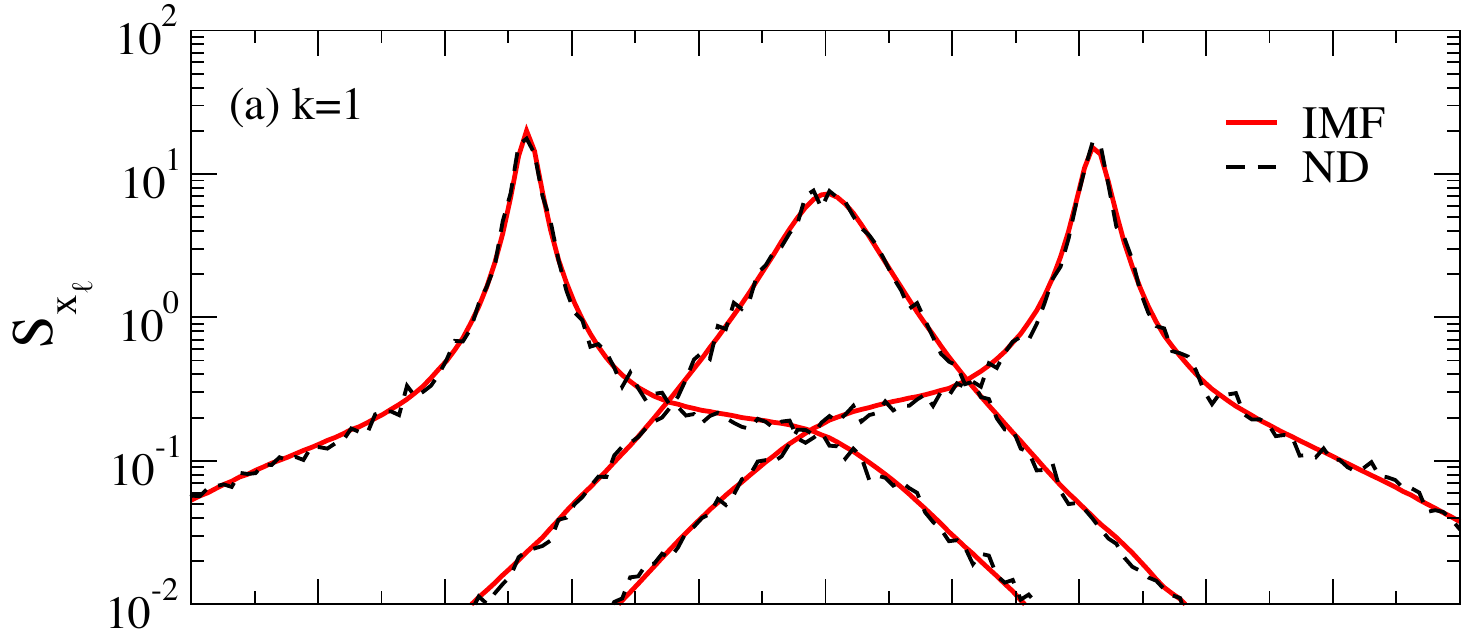}\\
\includegraphics[width=1\linewidth]{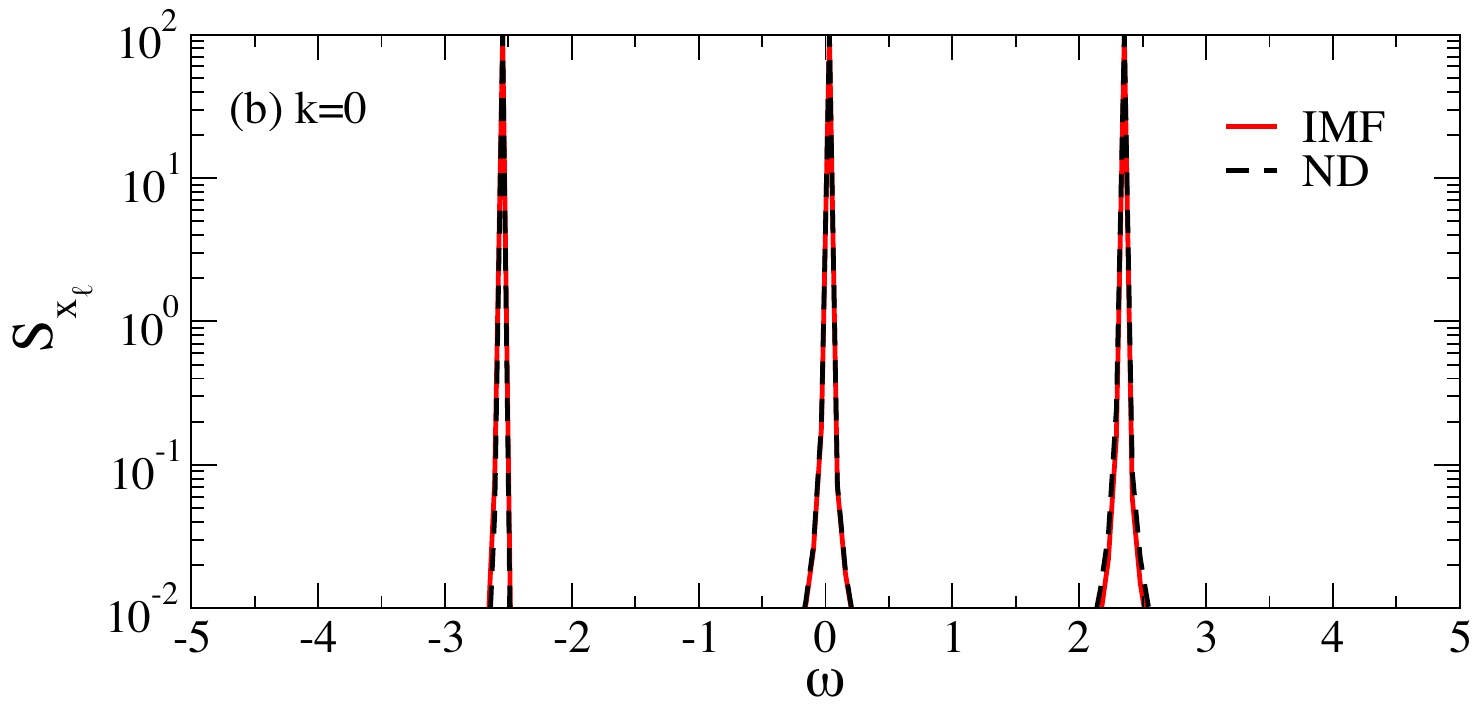}
\caption{\textbf{Single-oscillator dynamics with (a) and without (b) connectivity disorder.} Power spectra for three distinct oscillators with $\omega_6=-2.6, \omega_{32}=0.02, \omega_{79}=2.3$ and with Gaussian disorder in the connectivity ($k=1$ in (a)) or without disorder in the connectivity ($k=0$ in (b)). Red solid lines (IMF) and black dashed lines (ND) show matching results for both network types. The analysis is conducted with a frequency variability of $\sigma_\omega=1$, coupling strength $|K|=1$, in a system of size $N=10^4$ and over a time window $T=10^4$. All results were obtained from a single realization of frequency and connectivity disorder: $R=1, M=1$.} 
\label{fig:fig3}
\end{figure}

\begin{figure}
\centering 
\includegraphics[width=1\linewidth]{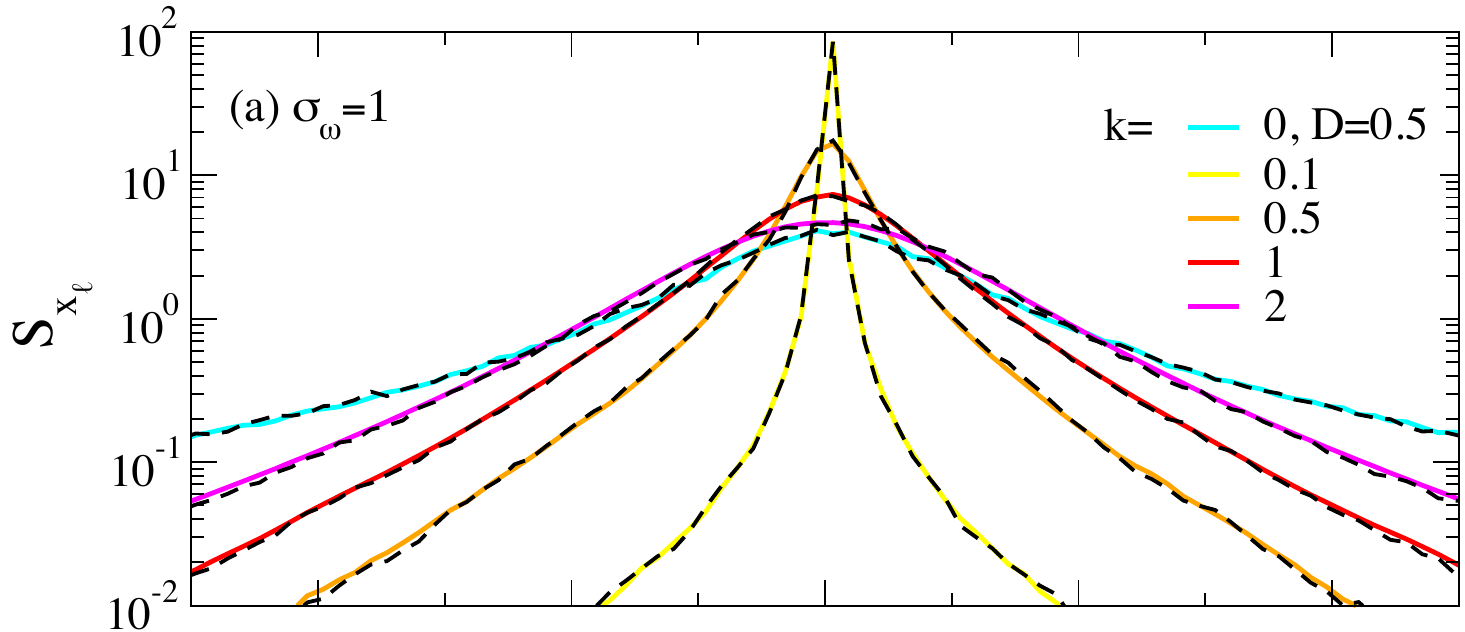}\\
\includegraphics[width=1\linewidth]{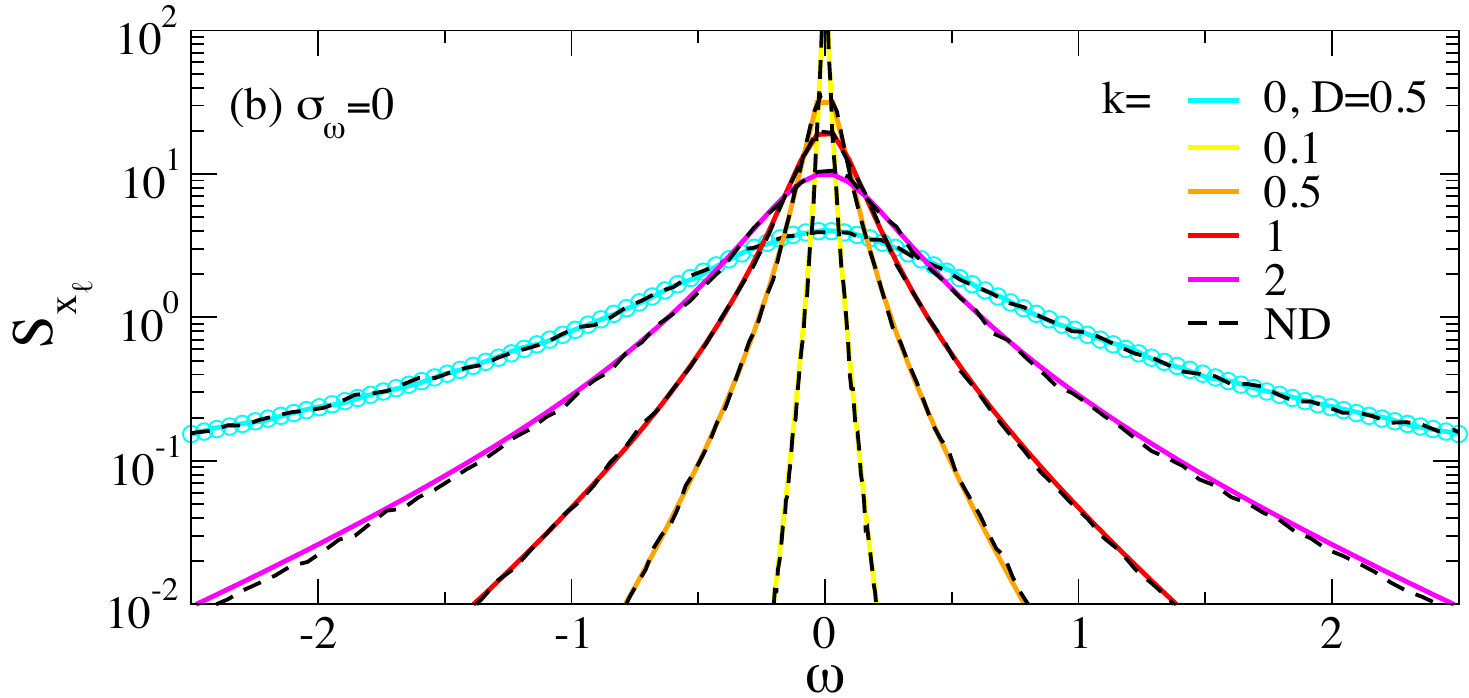}
\caption{\textbf{Dependence of single-oscillator power spectrum on network disorder with (a) and without (b) frequency variability}. Power spectrum of 32nd oscillator $S_{x_\ell}$ for increasing values of the Gaussian network disorder $k$ as indicated (a) with natural frequency $\omega_{32}=0.02$ (same as 
the middle curve in \bi{fig3})  and with a variability in natural frequencies of the network oscillators of $\sigma_\omega=1$. In (b), the same spectra but without frequency variability (all oscillators $\omega_\ell=0$). Dashed lines represent the full network dynamics (ND) for a single realization ($R=1$) with $N=10000$. Solid lines depict the IMF approach after iterations within $I=20$ in (a) and $I=50$ in (b), both with $M=1, N=10^4, T=10^5$. 
All data correspond to $D=0$, except the cyan line (IMF) and its corresponding black dashed line (ND) for $D=0.5$ in both panels and cyan circles (Eq.~\ref{eq:d}) in (b).
In (a), $|K|=1$; in (b), $|K|=0$. }
\label{fig:fig4}
\end{figure}

\begin{figure}
\centering
\includegraphics[width=1\linewidth]{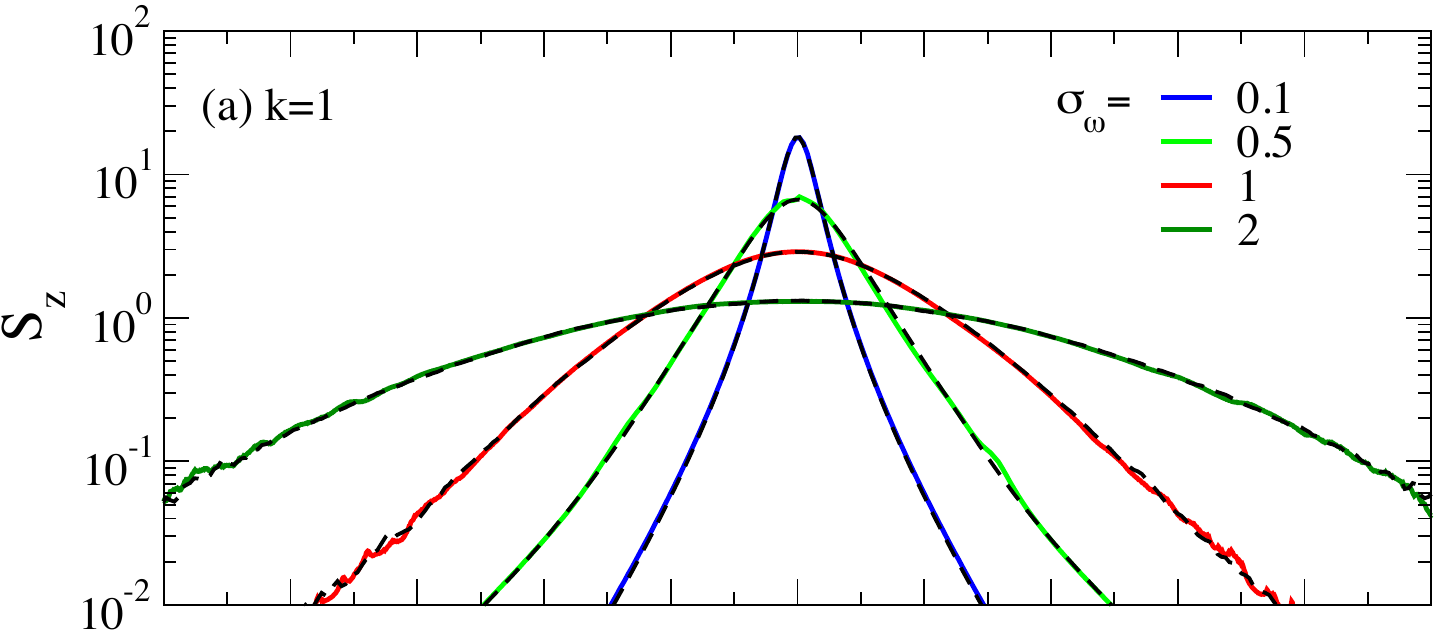}\\
\includegraphics[width=1\linewidth]{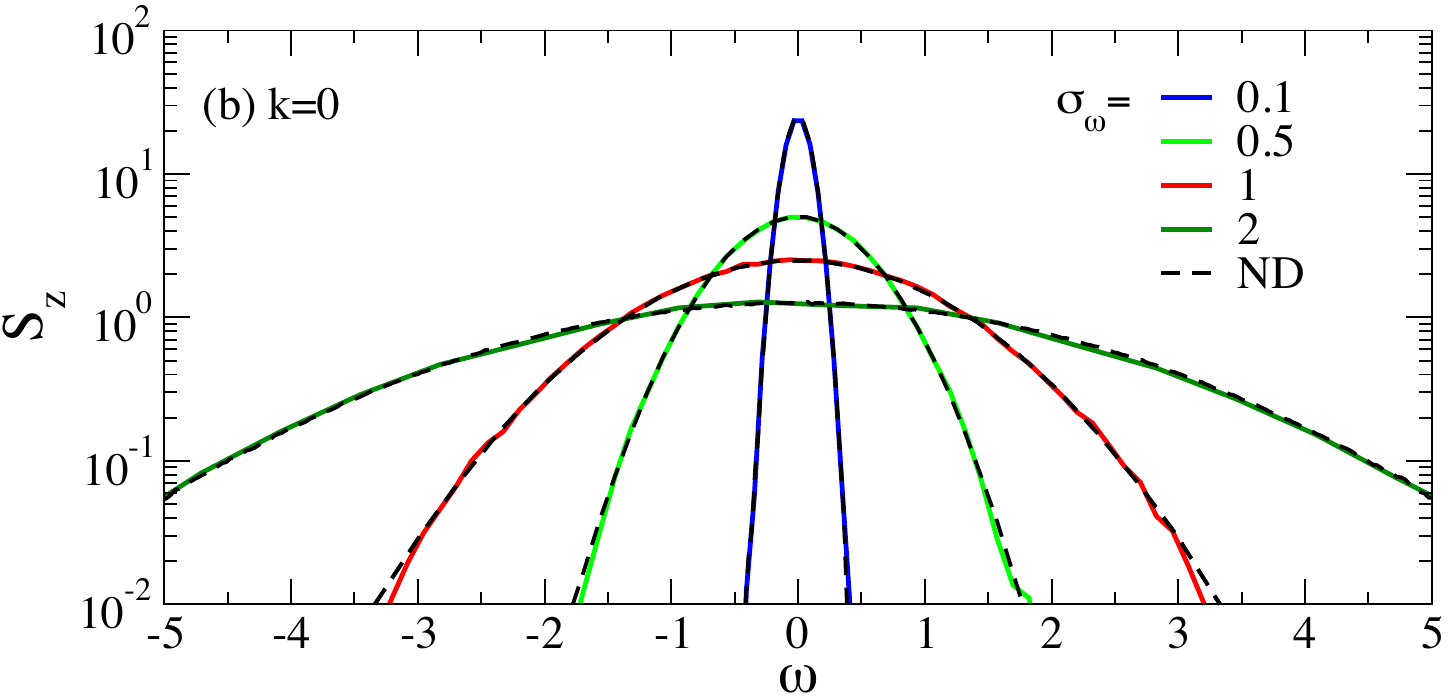}
\caption{\textbf{Average power spectra $S_z=\langle S_{x_\ell} \rangle_\ell$ with (a) and without (b) connectivity disorder}. Spectra averaged over all oscillators in the network with Gaussian connectivity ($k=1$ in (a)) or without ($k=0$ in (b)); in both panels we indicate increasing disorder in frequencies, $\sigma_\omega$. Dashed lines indicate ND with $N=10^3$. We used $R=10^3$ realizations with time window $T=10^5$ (a) and  $R=200$ realizations with time window $T=10^4$ (b). Solid lines illustrate the IMF method with $I=20$ iterations, employing $M=100$ and $T=10^5$. We set the parameter $K$ equal to the value of $|\sigma_\omega|$.}
\label{fig:fig5} 
\end{figure}

\begin{figure}
\centering
\includegraphics[width=1\linewidth]{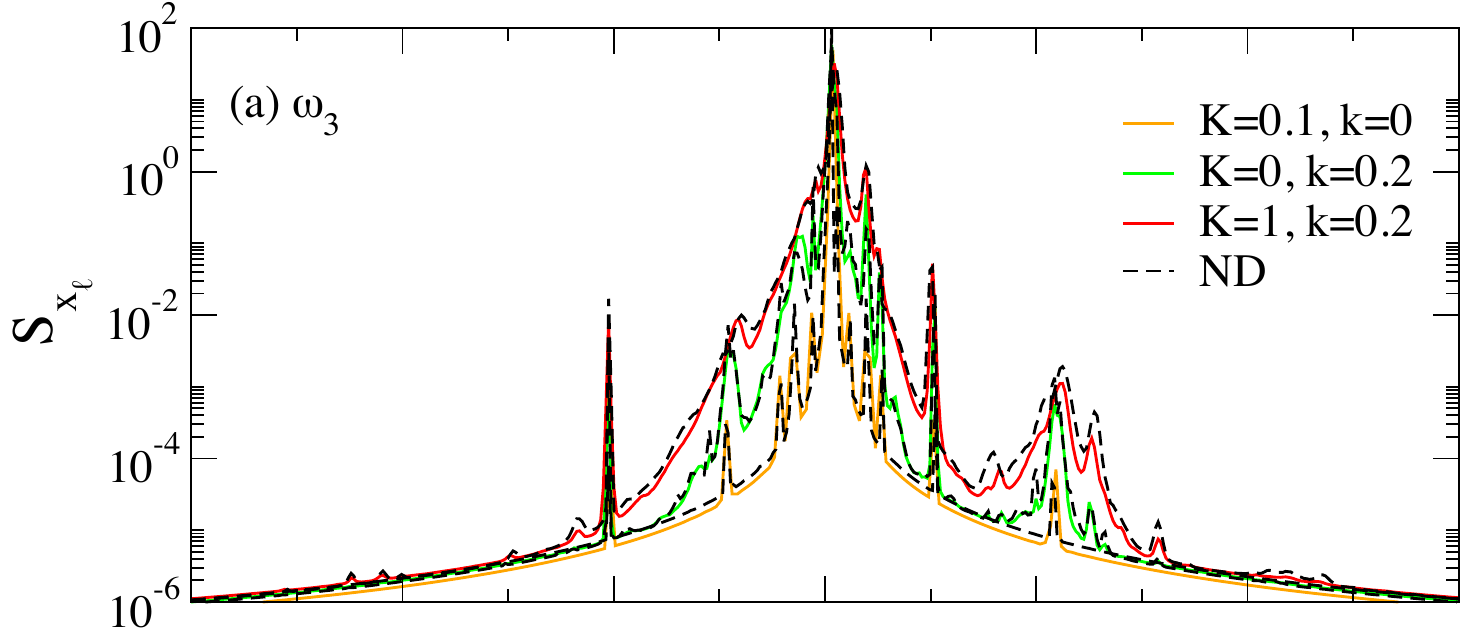}\\
\includegraphics[width=1\linewidth]{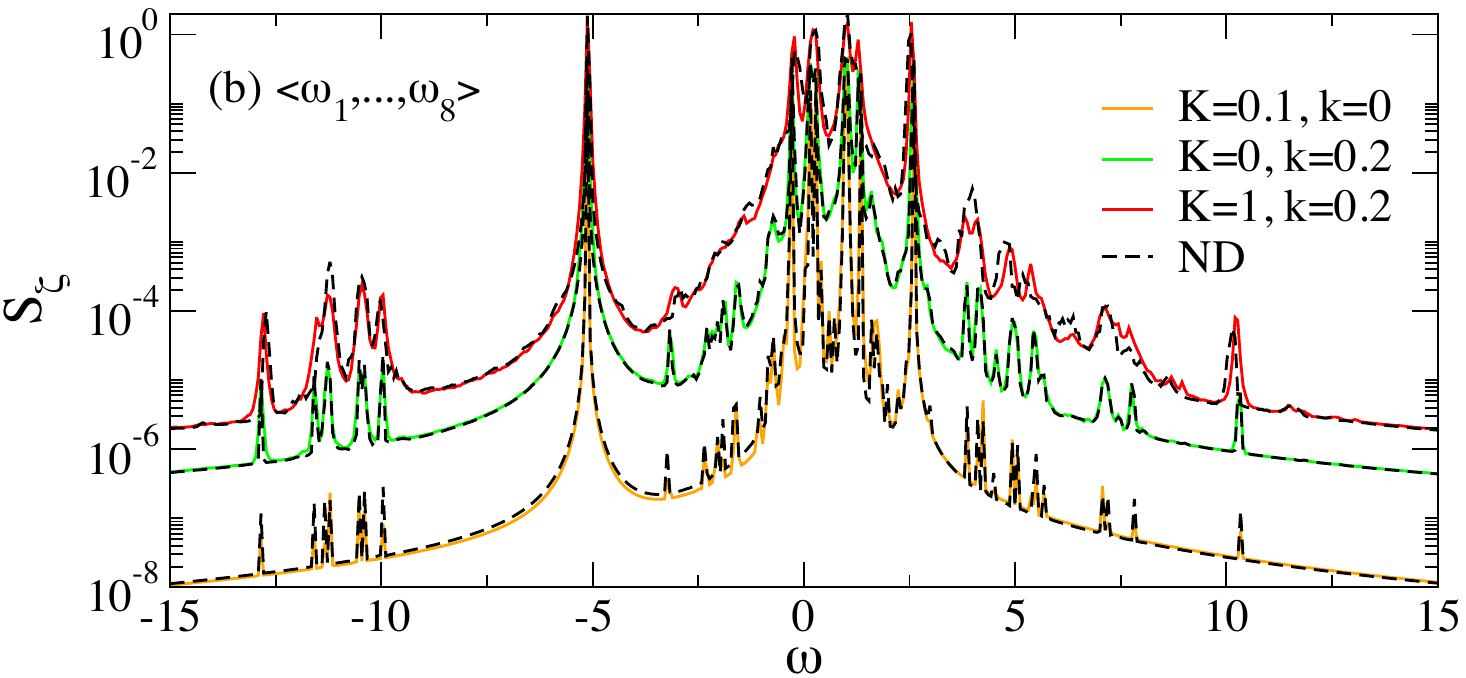} 
\caption{\textbf{Impact of mean coupling $K$ at finite network size $N$ on power spectra of single oscillator (a) and network noise (b)}. Spectrum of the third oscillator $S_{x_\ell}$ with $\omega_3=0.13$ in (a) and  noise spectrum $S_\zeta$ in (b) for a small network with $N=8$. Parameters: $\sigma_\omega=2$, $N=8$, $M=1000$, $R=1000$, $T=10^4$, frequencies $\omega=[0.97,0.26,0.13,2.6,1.3,-5.1,-0.31,1.04]$. IMF results in orange ($K=0.1, k=0$), green ($K=0, k=0.2$), and red ($K=1, k=0.2$) lines are compared with ND (dashed lines).} 
\label{fig:fig6}
\end{figure}

In the following, we conduct a comparative analysis of the power spectrum obtained using two distinct 
methods: (i) the iterative mean field (IMF) method, a stochastic mean field technique that simplifies the system of $N$ oscillators to a single effective 
oscillator, and (ii) the network dynamics (ND) approach, wherein we solve the differential equations for $N$ oscillators, \e{dottheta}. All the IMF results, depicted as colored lines, agree with the ND results, shown as dashed lines, confirming the validity of the mean-field approach. 

Figure \ref{fig:fig3} displays the spectra $S_{x_\ell}$  of individual oscillators, contrasting a Kuramoto model with disordered connectivity ($k=1$, panel a) with a uniformly connected one ($k=0$,  panel b), while the frequency variability $\sigma_\omega$ is set to 1 in both cases. We examine the spectra of three oscillators with distinct natural frequencies $\{\omega_{6}, \omega_{32}, \omega_{79}\}$ to show how they are differently affected by the network noise (the latter is proportional to $S_z$ shown in \bi{fig5} by the red curves).
For the spectra in \bi{fig3}(a), it is important to realize that the main share of the input network noise that shapes these spectra is around zero frequency. Hence the oscillator with a negative (positive) eigenfrequency displays a shoulder on the right (left) of its eigenfrequency - exactly in the frequency band around zero, whereas the oscillator with an eigenfrequency about zero does not display such an asymmetry. 
Without disorder in the network connectivity, \bi{fig3}(b), the oscillators' spectra become distinctly narrow with clear separation, in contrast to the overlapping profiles seen in the heterogeneous network model in \bi{fig3}(a).

The effect of network disorder becomes clearer when we increase $k$ in steps from 0.1 to 2, shown in \bi{fig4}(a). Here we maintain, for different values of $k$, the identical disorder realization of the connectivities $G_{\ell m}$ and of the frequencies $G_\ell$ as in \bi{fig3}; those values where defined in \e{omega} and \e{kij}. In \bi{fig4}, we focus on the 32nd oscillator, shown as the central curve with $\omega_{32}=0.02$ in \bi{fig3}, among a total of $N=10^4$ oscillators. As the strength of connectivity disorder, $k$, increases, we note a gradual broadening of the spectra alongside a reduction in peak amplitude, while the area beneath the curves remains constant; this simply reflects the fact that the variance of the pointer $x_\ell=e^{i\theta_\ell}$ is always one,  $\langle |x_\ell|^2 \rangle =1$. Instead of network disorder,  intrinsic white noise (here with $D=0.5$) can also broaden the peak.

Going from \bi{fig4}(a) to (b), we examine  the impact  of removing frequency variability on the power spectra. We observe that the spectrum becomes more narrow and its amplitude increases. However, the narrowing of the peaks is only moderate. 
Therefore, the peak width seems to be mainly determined by the heterogeneity of the connectivity. 
While we consider $N=10^4$ oscillators for both \bi{fig3} and \bi{fig4}, we obtain similar results for $N=10^3$ oscillators, confirming that the outcomes are independent of $N$ as long as $K^2/N \ll k^2$ (otherwise the network noise \e{zeta1} will depend on $N$).

In \bi{fig5}, we turn to the network averaged spectra $S_z$ for different frequency disorder from $\sigma_\omega=0.1$ to 2 in the presence and absence of connectivity disorder, $k=1$ in (a) and $k=0$ in (b), respectively. The red curves for $\sigma_\omega=1$ correspond to the network-averaged spectra of \bi{fig3}. We recall that there is a simple relationship between the network noise and the network averaged spectra, \e{zeta1}, according to which both are proportional to each other.
Put differently, in \bi{fig5} we look at scaled versions of the network noise power spectra $S_\zeta$. We again emphasize that the IMF method (colored solid lines) yields excellent agreement in reproducing the spectra of the true network noise obtained from ND simulations (dashed lines).

In the absence of connectivity disorder, the spectra in \bi{fig5}(b) attain the shape of a Gaussian, $S_z=\sqrt{2\pi} e^{-\omega^2/2\sigma_\omega^2}/\sigma_\omega$. This is plausible for $K=0$ where in the network noise we add up deterministic oscillators with randomly drawn eigenfrequencies according to a Gaussian distribution. It remains a valid approximation though for values of $K$ below the critical value $K_c$. We emphasize that in the presence of connectivity disorder \bi{fig5}(a) the shape of the spectra is neither captured by a Lorentzian nor by a Gaussian.

So far, we have concentrated on large networks with $N\gg1$, where finite-size corrections are negligible. One may ask whether the theory holds equally well for smaller networks, and if the contribution that stems from the mean connection strength in \e{spec_zeta} becomes relevant in this case. In Figure \ref{fig:fig6}, we analyze a system comprising $N=8$ oscillators with fixed natural frequencies $\omega_\ell$, $\ell=1,\ldots,8$, and investigate the impact of varying both the average coupling constant $K$ and the coupling disorder strength $k$. In panel (a), we plot the power spectrum of the third oscillator for a finite network disorder ($k=0.2$) for vanishing mean connectivity ($K=0$, green line) and for a finite but subcritical value of the mean connectivity ($K=1<K_c$, red line); clearly, the value of the mean connectivity has an impact on the spectrum (green and red lines differ substantially). This is due to the finite size of the network because $K$ enters the mean-field theory only by a factor of $K^2/N$ that vanishes in the thermodynamic limit. For the small number of oscillators considered here, the power spectrum has a complicated shape with several distinct peaks, the dominating one being located at the eigenfrequency of the oscillator, $\omega_3=0.13$. The spectrum is furthermore shaped by the input noise that is illustrated in panel (b); to determine the network noise, we average here over different realizations of the connectivity disorder. For both the oscillator and noise spectra, the agreement between the stochastic mean-field IMF method and the ND simulations is reasonably good; in particular, the alignment for the oscillator spectra confirms the finite-size correction of our theory.  Our results clearly indicate the necessity of finite size correction when the condition $N\gg1$ is not met.  We also demonstrate that our theory does not need a nonvanishing value of the connectivity disorder but also works for the original Kuramoto model that has only frequency disorder ($k=0, K=0.1$, orange line); also here the IMF result agrees fairly well with the ND simulations.

One might wonder how mean field methods yield accurate results for systems as small as comprising $N=8$ oscillators. While we fixed the natural frequencies for all $N=8$ oscillators in the networks analyzed for Fig.~\ref{fig:fig6}, we averaged over multiple instantiations of the connectivity matrix $K_{\ell m}$, which allows us to obtain ensemble-averaged observables that remain well described by the IMF approach. In the case of large system sizes $N\gg1$, the network dynamics becomes self-averaging, and the IMF approach yields correct results for observables that are determined using ND for networks with a single instantiation of the connectivity disorder (see Figs.~\ref{fig:fig3}-\ref{fig:fig4}).

In our study, for large system sizes ($N\gg1$), we observe a notable insensitivity of the system dynamics to the coupling strength $K$ when $|K| < K_c$, i.e., below the onset of synchronization.  Specifically, the resulting power spectra do not depend on the value of $K$, including negative ones; for instance, spectra for $K=0, -\sigma_\omega, +\sigma_\omega$ agree.  

\section{Summary and Conclusions}
In this study, we have explored the asynchronous states of the Kuramoto model with a specific focus on the power spectra of individual oscillators. We comprehensively analyzed how two distinct types of disorder---variations in natural frequencies and network connectivities---uniquely affect the dynamics of single oscillators. This approach allows us to 
understand the model dynamics in a variety of coupling conditions, including zero or non-vanishing average coupling constants as well as homogeneous or heterogeneous network structures, hence various situations that have previously been studied in the literature \cite{PhysRevLett.106.054102,PhysRevE.89.012910,CUMIN2007181,Choi2019}. 
Building upon the groundwork laid by Stiller and Radons \cite{stiller}, we have further developed the mean-field method to account for the finite average coupling and also rederived the framework in a simplified manner. The stochastic iterative method applies to the homogeneous network setup---the classical Kuramoto model---and also provides an additional finite-size correction. In both homogeneous and heterogeneous networks, we have successfully reproduced the power spectra of large network simulations by employing this stochastic iterative method.

Our analysis reveals that all spectra exhibit a significant decrease in peak height and a concomitant broadening as we transition from homogeneous ($k=0$) to heterogeneous ($k=1$) connectivities (see \bi{fig3}). The variation in $k$ increases the network noise and expands the range of frequencies. Put differently, the static disorder in the connectivity translates into a dynamic network noise. This mirrors findings in physical and biological systems, where heterogeneity in interactions often leads to more noisy behavior \cite{BOCCALETTI2006175,Buzsaki2014}. In ecological and social networks, for instance, the introduction of diverse interaction strengths or patterns often leads to a richer array of system states and behaviors, reflecting a balance between resilience and flexibility \cite{May1972,Barabasi99}.

The large and small networks with heterogeneous connections in our study include both positive and negative coupling coefficients $K_{\ell m}$. As we ensure the average coupling is in the range from $-K_c$ to $K_c$, the system is always poised in an asynchronous state. For large systems $N \gg 1$, we observe a notable insensitivity of the system dynamics to the average coupling strength $K$ when  $|K| < K_c$. However, this observation for large systems contrasts sharply with the dynamics observed in smaller systems, such as $N = 8$, where the specific value of  $K$  (while still  $|K| < K_c$) becomes significant, distinctly influencing the system's behavior as shown in \bi{fig6}. Interestingly, for $N= 8$, there still exists a symmetry in the system's response to positive and negative values of  $K$  of equal magnitude; choosing  $K = 1$ or  $K = -1$ yields identical spectra. The existence of both positive and negative couplings prevents excessive synchronization in neural networks \cite{PhysRevLett.106.054102} and maintains species diversity in ecological and social models \cite{Girn2016}. 

There are several extensions of the model that come to mind, which could be treated with a similar stochastic mean-field approach as applied in this paper.
For instance, endowing the phase oscillators with inertia provides a better network model in many situations, e.g. for power grids \cite{PhysRevE.109.024212}. Furthermore, it is of interest in many fields, for example in neuroscience, to explore how the network's oscillators respond to external perturbations (in the neural context this could be a sensory signal to be represented in the oscillators' activity). The iterative mean field method can certainly be generalized to calculate the self-consistent response functions (susceptibilities) of the network. Lastly, we could also consider with the similar method a network of networks, as it has been done previously for simple rotator networks \cite{RanLin23}.

A remaining open problem is the analytical solution of the self-consistent network noise statistics. For simple rotators, this can be done by solving simple differential equations \cite{VanLin18,RanLin22}. For the Kuramoto model, the resulting noise term is multiplicative and non-Gaussian (the noise $\zeta(t)$ \emph{is} Gaussian but the last term in \e{thetadotmean} is not). Here a new approach is needed. So, there are many exciting problems left for future research to better understand the asynchronous state of the Kuramoto model.

\bibliography{KM.bib}
\end{document}